\documentclass[submission]{eptcs}

\usepackage{proof}
\usepackage{xspace}
\usepackage{fixltx2e} % stops large figures from floating out-of-order
\usepackage{xcolor}
\usepackage{amsmath}
\usepackage{amssymb}
\usepackage{amsthm}
\usepackage{xifthen}
\usepackage{caption}
\usepackage[]{subcaption}
\usepackage{hyperref} % Provides the \url macros.
\usepackage{tikz}
\usetikzlibrary{automata}
\usepackage{cleveref}

\makeatletter
\renewcommand{\infer}{\@ifnextchar *{\@inferSteps}{\@inferOneStep}}
\makeatother

\newcommand{\lltens}{\mathbin{\otimes}}
\newcommand{\llone}{\mathbf{1}}
\newcommand{\llplus}{\mathbin{\oplus}}
\newcommand{\llzero}{\mathbf{0}}
\newcommand{\llpar}{\mathbin{\wp}}
\newcommand{\llwith}{\mathbin{\&}}

\newcommand{\false}{f}
\newcommand{\true}{t}

\newcommand{\pand}{\mathbin{\wedge\kern-1.5pt^+}}
\newcommand{\ptrue}{\true^+}
\newcommand{\por}{\mathbin{\vee}}
\newcommand{\bigpor}{\bigvee}
\newcommand{\pfalse}{\false^+}
\newcommand{\nimp}{\mathbin{\supset}}
\newcommand{\nfalse}{\false^-}
\newcommand{\nand}{\mathbin{\wedge\kern-1.5pt^-}}
\newcommand{\ntrue}{\true^-}

\newcommand{\certificate}[1]{\textsf{#1}}
\newcommand{\stopcert}{\certificate{stop}}
\newcommand{\synccert}{\certificate{sync}}
\newcommand{\asynccert}{\certificate{async}}
\newcommand{\bipolecert}[1][]{%
  \certificate{bipole}%
  \ifthenelse{\isempty{#1}}{}{$_{#1}$}%
}
\newcommand{\deccert}{\certificate{decproc}}
\newcommand{\invcert}{\certificate{inv}}
\newcommand{\coinvcert}{\certificate{co\text{-}inv}}

\newcommand{\eqsc}{{=^s_c}}
\newcommand{\neqfc}{{\neq^f_c}}
\newcommand{\ptruec}{{\ptrue_c}}
\newcommand{\nfalsec}{{\nfalse_c}}
\newcommand{\pandc}{{\pand_c}}
\newcommand{\nimpc}{{\nimp_c}}
\newcommand{\eqfc}{{=^f_c}}
\newcommand{\neqsc}{{\neq^s_c}}
\newcommand{\pfalsec}{{\pfalse_c}}
\newcommand{\ntruec}{{\ntrue_c}}
\newcommand{\porc}{{\por_c}}
\newcommand{\nandc}{{\nand_c}}
\newcommand{\existsc}{{\exists_c}}
\newcommand{\forallc}{{\forall_c}}
\newcommand{\neqfe}{{\neq^f_e}}
\newcommand{\eqse}{{=^s_e}}
\newcommand{\nfalsee}{{\nfalse_e}}
\newcommand{\ptruee}{{\ptrue_e}}
\newcommand{\nimpe}{{\nimp_e}}
\newcommand{\pande}{{\pand_e}}
\newcommand{\nande}{{\nand_e}}
\newcommand{\pore}{{\por_e}}
\newcommand{\foralle}{{\forall_e}}
\newcommand{\existse}{{\exists_e}}
\newcommand{\storelc}{{\text{store}_L}}
\newcommand{\storerc}{{\text{store}_R}}
\newcommand{\decidele}{{\text{decide}_L}}
\newcommand{\decidere}{{\text{decide}_R}}
\newcommand{\releasele}{{\text{release}_L}}
\newcommand{\releasere}{{\text{release}_R}}
\newcommand{\inde}{{\text{ind}}}
\newcommand{\coinde}{{\text{co-ind}}}
\newcommand{\mule}{{\mu\text{-unfold}_L}}
\newcommand{\nure}{{\nu\text{-unfold}_R}}
\newcommand{\nule}{{\nu\text{-unfold}_L}}
\newcommand{\mure}{{\mu\text{-unfold}_R}}

\newcommand{\Nscr}{\mathcal{N}}
\newcommand{\Pscr}{\mathcal{P}}
\newcommand{\cert}[1]{#1}
\newcommand{\premise}[1]{&#1}
\newcommand{\tabl}[1]{#1}

\newcommand{\syncR}[4][]{{%
  \ifthenelse{\isempty{#1}}{}{\cert{{#1}\;\!{:}\,}}% certificate
  \ifthenelse{\isempty{#2}}{}{\tabl{{#2}\;\!{;}\,}}{}% empty left-hand side
  \vdash%
  {#3}\Downarrow{}\ifthenelse{\isempty{#4}}{}{\tabl{\,{;}\;\!{#4}}}% right-hand side
}}
\newcommand{\syncL}[4][]{{%
  \ifthenelse{\isempty{#1}}{}{\cert{{#1}\;\!{:}\,}}% certificate
  \ifthenelse{\isempty{#2}}{}{\tabl{{#2}\;\!{;}\,}}{}\Downarrow{#3}% left-hand side
  \vdash%
  {}\ifthenelse{\isempty{#4}}{}{\tabl{\,{;}\;\!{#4}}}% empty right-hand side
}}
\newcommand{\async}[7][]{{%
  \ifthenelse{\isempty{#1}}{}{\cert{{#1}\;\!{:}\,}}% certificate
  \ifthenelse{\isempty{#2}}{}{\tabl{{#2}\;\!{;}\,}}{#3}\Uparrow{#4}% left-hand side
  \vdash%
  {#5}\Uparrow{#6}\ifthenelse{\isempty{#7}}{}{\tabl{\,{;}\;\!{#7}}}% right-hand side
}}

\newcommand{\certcolour}[1]{%
  \ifthenelse{\isempty{#1}}{%
    \expandafter\renewcommand{\cert}[1]{}%
    \expandafter\renewcommand{\premise}[1]{}%
  }{\expandafter%
    \expandafter\renewcommand{\cert}[1]{{\color{#1}##1}}%
    \expandafter\renewcommand{\premise}[1]{&{\color{#1}##1}}%
  }%
}
\newcommand{\tablcolour}[1]{%
  \ifthenelse{\isempty{#1}}{%
    \expandafter\renewcommand{\tabl}[1]{}%
  }{\expandafter%
    \expandafter\renewcommand{\tabl}[1]{{\color{#1}##1}}%
  }%
}

\definecolor{thecertcolour}{hsb}{0.58,1,1}
\definecolor{thetablcolour}{hsb}{0.05,1,1}

\certcolour{thecertcolour}
\tablcolour{}

\newcommand{\eqsl}{
  \infer[{=^s_L}\dag]
    {\async[\Xi_0]{\Theta}{\Nscr}{s=t,\Gamma}{\Delta}{}{\Upsilon}}
    {\async[\Xi_1\theta]{\Theta}{\Nscr\theta}{\Gamma\theta}{\Delta\theta}{}{\Upsilon}
     \premise{\eqsc(\Xi_0,\Xi_1)}}
}
\newcommand{\neqfr}{
  \infer[{\neq^f_R}\dag]
    {\async[\Xi_0]{\Theta}{\Nscr}{}{s\neq{}t}{}{\Upsilon}}
    {\async[\Xi_1\theta]{\Theta}{\Nscr\theta}{}{}{}{\Upsilon}
     \premise{\neqfc(\Xi_0,\Xi_1)}}
}
\newcommand{\ptruel}{
  \infer[{\ptrue_L}]
    {\async[\Xi_0]{\Theta}{\Nscr}{\ptrue,\Gamma}{\Delta}{}{\Upsilon}}
    {\async[\Xi_1]{\Theta}{\Nscr}{\Gamma}{\Delta}{}{\Upsilon}
     \premise{\ptruec(\Xi_0,\Xi_1)}}
}
\newcommand{\nfalser}{
  \infer[{\nfalse_R}]
    {\async[\Xi_0]{\Theta}{\Nscr}{}{\nfalse}{}{\Upsilon}}
    {\async[\Xi_1]{\Theta}{\Nscr}{}{}{}{\Upsilon}
     \premise{\nfalsec(\Xi_0,\Xi_1)}}
}
\newcommand{\pandl}{
  \infer[{\pand_L}]
    {\async[\Xi_0]{\Theta}{\Nscr}{A_1\pand{}A_2,\Gamma}{\Delta}{}{\Upsilon}}
    {\async[\Xi_1]{\Theta}{\Nscr}{A_1,A_2,\Gamma}{\Delta}{}{\Upsilon}
     \premise{\pandc(\Xi_0,\Xi_1)}}
}
\newcommand{\nimpr}{
  \infer[{\nimp_R}]
    {\async[\Xi_0]{\Theta}{\Nscr}{}{A_1\nimp{}A_2}{}{\Upsilon}}
    {\async[\Xi_1]{\Theta}{\Nscr}{A_1}{A_2}{}{\Upsilon}
     \premise{\nimpc(\Xi_0,\Xi_1)}}
}
\newcommand{\eqfl}{
  \infer[{=^f_L}\ddag]
    {\async[\Xi_0]{\Theta}{\Nscr}{s=t,\Gamma}{\Delta}{}{\Upsilon}}
    {\premise{\eqfc(\Xi_0)}}
}
\newcommand{\neqsr}{
  \infer[{\neq^s_R}\ddag]
    {\async[\Xi_0]{\Theta}{\Nscr}{}{s\neq{}t}{}{\Upsilon}}
    {\premise{\neqsc(\Xi_0)}}
}
\newcommand{\pfalsel}{
  \infer[{\pfalse_L}]
    {\async[\Xi_0]{\Theta}{\Nscr}{\pfalse,\Gamma}{\Delta}{}{\Upsilon}}
    {\premise{\pfalsec(\Xi_0)}}
}
\newcommand{\ntruer}{
  \infer[{\ntrue_R}]
    {\async[\Xi_0]{\Theta}{\Nscr}{}{\ntrue}{}{\Upsilon}}
    {\premise{\ntruec(\Xi_0)}}
}
\newcommand{\porl}{
  \infer[{\por_L}]
    {\async[\Xi_0]{\Theta}{\Nscr}{A_1\por{}A_2,\Gamma}{\Delta}{}{\Upsilon}}
    {\async[\Xi_1]{\Theta}{\Nscr}{A_1,\Gamma}{\Delta}{}{\Upsilon} &
     \async[\Xi_2]{\Theta}{\Nscr}{A_2,\Gamma}{\Delta}{}{\Upsilon}
     \premise{\porc(\Xi_0,\Xi_1,\Xi_2)}}
}
\newcommand{\nandr}{
  \infer[{\nand_R}]
    {\async[\Xi_0]{\Theta}{\Nscr}{}{A_1\nand{}A_2}{}{\Upsilon}}
    {\async[\Xi_1]{\Theta}{\Nscr}{}{A_1}{}{\Upsilon} &
     \async[\Xi_2]{\Theta}{\Nscr}{}{A_2}{}{\Upsilon}
     \premise{\nandc(\Xi_0,\Xi_1,\Xi_2)}}
}
\newcommand{\existsl}{
  \infer[{\exists_L}]
    {\async[\Xi_0]{\Theta}{\Nscr}{\exists{}x.\,C\,x,\Gamma}{\Delta}{}{\Upsilon}}
    {\async[\Xi_1\,y]{\Theta}{\Nscr}{C\,y,\Gamma}{\Delta}{}{\Upsilon}
     \premise{\existsc(\Xi_0,\Xi_1)}}
}
\newcommand{\forallr}{
  \infer[{\forall_R}]
    {\async[\Xi_0]{\Theta}{\Nscr}{}{\forall{}x.\,C\,x}{}{\Upsilon}}
    {\async[\Xi_1\,y]{\Theta}{\Nscr}{}{C\,y}{}{\Upsilon}
     \premise{\forallc(\Xi_0,\Xi_1)}}
}

\newcommand{\neqfl}{
  \infer[{\neq^f_L}]
    {\syncL[\Xi_0]{\Theta}{t\neq{}t}{\Upsilon}}
    {\premise{\neqfe(\Xi_0)}}
}
\newcommand{\eqsr}{
  \infer[{=^s_R}]
    {\syncR[\Xi_0]{\Theta}{t=t}{\Upsilon}}
    {\premise{\eqse(\Xi_0)}}
}
\newcommand{\nfalsel}{
  \infer[{\nfalse_L}]
    {\syncL[\Xi_0]{\Theta}{\nfalse}{\Upsilon}}
    {\premise{\nfalsee(\Xi_0)}}
}
\newcommand{\ptruer}{
  \infer[{\ptrue_R}]
    {\syncR[\Xi_0]{\Theta}{\ptrue}{\Upsilon}}
    {\premise{\ptruee(\Xi_0)}}
}
\newcommand{\nimpl}{
  \infer[{\nimp_L}]
    {\syncL[\Xi_0]{\Theta}{A_1\nimp{}A_2}{\Upsilon}}
    {\syncR[\Xi_1]{\Theta}{A_1}{\Upsilon} &
     \syncL[\Xi_2]{\Theta}{A_2}{\Upsilon}
     \premise{\nimpe(\Xi_0,\Xi_1,\Xi_2)}}
}
\newcommand{\pandr}{
  \infer[{\pand_R}]
    {\syncR[\Xi_0]{\Theta}{A_1\pand{}A_2}{\Upsilon}}
    {\syncR[\Xi_1]{\Theta}{A_1}{\Upsilon} &
     \syncR[\Xi_2]{\Theta}{A_2}{\Upsilon}
     \premise{\pande(\Xi_0,\Xi_1,\Xi_2)}}
}
\newcommand{\nandl}{
  \infer[{\nand_L}]
    {\syncL[\Xi_0]{\Theta}{A_1\nand{}A_2}{\Upsilon}}
    {\syncL[\Xi_1]{\Theta}{A_i}{\Upsilon}
     \premise{\nande(\Xi_0,\Xi_1,i)}}
}
\newcommand{\porr}{
  \infer[{\por_R}]
    {\syncR[\Xi_0]{\Theta}{A_1\por{}A_2}{\Upsilon}}
    {\syncR[\Xi_1]{\Theta}{A_i}{\Upsilon}
     \premise{\pore(\Xi_0,\Xi_1,i)}}
}
\newcommand{\foralll}{
  \infer[{\forall_L}]
    {\syncL[\Xi_0]{\Theta}{\forall{}x.\,C\,x}{\Upsilon}}
    {\syncL[\Xi_1]{\Theta}{C\,t}{\Upsilon}
     \premise{\foralle(\Xi_0,\Xi_1,t)}}
}
\newcommand{\existsr}{
  \infer[{\exists_R}]
    {\syncR[\Xi_0]{\Theta}{\exists{}x.\,C\,x}{\Upsilon}}
    {\syncR[\Xi_1]{\Theta}{C\,t}{\Upsilon}
     \premise{\existse(\Xi_0,\Xi_1,t)}}
}

\newcommand{\storel}{
  \infer[{\text{S}_L}]
    {\async[\Xi_0]{\Theta}{}{N,\Gamma}{\Delta}{}{\Upsilon}}
    {\async[\Xi_1]{\Theta}{N}{\Gamma}{\Delta}{}{\Upsilon}
     \premise{\storelc(\Xi_0,\Xi_1)}}
}
\newcommand{\storer}{
  \infer[{\text{S}_R}]
    {\async[\Xi_0]{\Theta}{}{}{P}{}{\Upsilon}}
    {\async[\Xi_1]{\Theta}{}{}{}{P}{\Upsilon}
     \premise{\storerc(\Xi_0,\Xi_1)}}
}
\newcommand{\decidel}{
  \infer[{\text{D}_L}]
    {\async[\Xi_0]{\Theta}{N}{}{}{}{\Upsilon}}
    {\syncL[\Xi_1]{\Theta}{N}{\Upsilon}
     \premise{\decidele(\Xi_0,\Xi_1)}}
}
\newcommand{\decider}{
  \infer[{\text{D}_R}]
    {\async[\Xi_0]{\Theta}{}{}{}{P}{\Upsilon}}
    {\syncR[\Xi_1]{\Theta}{P}{\Upsilon}
     \premise{\decidere(\Xi_0,\Xi_1)}}
}
\newcommand{\releasel}{
  \infer[{\text{R}_L}]
    {\syncL[\Xi_0]{\Theta}{P}{\Upsilon}}
    {\async[\Xi_1]{\Theta}{}{P}{}{}{\Upsilon}
     \premise{\releasele(\Xi_0,\Xi_1)}}
}
\newcommand{\releaser}{
  \infer[{\text{R}_R}]
    {\syncR[\Xi_0]{\Theta}{N}{\Upsilon}}
    {\async[\Xi_1]{\Theta}{}{}{N}{}{\Upsilon}
     \premise{\releasere(\Xi_0,\Xi_1)}}
}

\newcommand{\ind}{
  \infer[{\mu}]
    {\async[\Xi_0]{\Theta}{\Nscr}{\mu{}B\,\t,\Gamma}{\Delta}{}{\Upsilon}}
    {\async[\Xi_1\,\y]{\Theta}{}{B\,\Sscr\,\y}{\Sscr\,\y}{}{\Upsilon} &
     \async[\Xi_2]{\Theta}{\Nscr}{\Sscr\,\t,\Gamma}{\Delta}{}{\Upsilon}
     \premise{\inde(\Xi_0,\Xi_1,\Xi_2,\Sscr)}}
}
\newcommand{\coind}{
  \infer[{\nu}]
    {\async[\Xi_0]{\Theta}{\Nscr}{}{\nu{}B\,\t}{}{\Upsilon}}
    {\async[\Xi_1]{\Theta}{\Nscr}{}{\Sscr\,\t}{}{\Upsilon} &
     \async[\Xi_2\,\y]{\Theta}{}{\Sscr\,\y}{B\,\Sscr\,\y}{}{\Upsilon}
     \premise{\coinde(\Xi_0,\Xi_1,\Xi_2,\Sscr)}}
}
\newcommand{\mul}{
  \infer[{\mu_L}]
    {\async[\Xi_0]{\Theta}{\Nscr}{\mu{}B\,\t,\Gamma}{\Delta}{}{\Upsilon}}
    {\async[\Xi_1]{\Theta}{\Nscr}{B(\mu{}B)\t,\Gamma}{\Delta}{}{\Upsilon}
     \premise{\mule(\Xi_0,\Xi_1)}}
}
\newcommand{\nur}{
  \infer[{\nu_R}]
    {\async[\Xi_0]{\Theta}{\Nscr}{}{\nu{}B\,\t}{}{\Upsilon}}
    {\async[\Xi_1]{\Theta}{\Nscr}{}{B(\nu{}B)\t}{}{\Upsilon}
     \premise{\nure(\Xi_0,\Xi_1)}}
}
\newcommand{\nul}{
  \infer[{\nu_L}]
    {\syncL[\Xi_0]{\Theta}{\nu{}B\,\t}{\Upsilon}}
    {\syncL[\Xi_1]{\Theta}{B(\nu{}B)\t}{\Upsilon}
     \premise{\nule(\Xi_0,\Xi_1)}}
}
\newcommand{\mur}{
  \infer[{\mu_R}]
    {\syncR[\Xi_0]{\Theta}{\mu{}B\,\t}{\Upsilon}}
    {\syncR[\Xi_1]{\Theta}{B(\mu{}B)\t}{\Upsilon}
     \premise{\mure(\Xi_0,\Xi_1)}}
}

\newcommand{\mumall}{\ensuremath{\mu\text{MALL}}}
\newcommand{\mumallf}{\ensuremath{\mu\text{MALLF}}}
\newcommand{\muF}{\ensuremath{\mu{}F}}
\newcommand{\muFa}{\ensuremath{\mu{}F^a}}

\newcommand{\diam}[1]{\langle #1 \rangle}
\newcommand{\bisimv}{\hbox{\sl bisim}}
\newcommand{\bisim}[2]{\bisimv(#1,#2)}
\newcommand{\ssimv}{\hbox{\sl sim}}
\newcommand{\ssim}[2]{\ssimv(#1,#2)}
\newcommand{\ppathv}{\hbox{\sl path}}
\newcommand{\ppath}[2]{\ppathv(#1,#2)}
\newcommand{\act}{\hbox{\sl Act}\xspace}
\newcommand{\atrue}{\hbox{\sl true}\xspace}
\newcommand{\one}[3]{{#1}\buildrel{#2}\over\longrightarrow{#3}}
\newcommand{\tup}[1]{\langle #1 \rangle}

\newcommand{\switchable}{switchable}

\newcommand{\x}{\bar{x}}
\newcommand{\y}{\bar{y}}
\renewcommand{\t}{\bar{t}}

\newcommand\ie{\textit{i.e.}\xspace}

\newcommand\eg{\textit{e.g.}}
\newcommand\etc{\textit{etc.}\xspace}

\newtheorem{theorem}{Theorem}
\newtheorem{definition}{Definition}
\newtheorem{example}[theorem]{Example}

\newcommand{\Sscr}{S}

\title{A framework for proof certificates in finite state exploration}
\newcommand{\titlerunning}{FPC for finite state exploration}
\newcommand{\authorrunning}{Q. Heath and D. Miller}

\begin{document}
\author{Quentin Heath and Dale Miller
        \institute{Inria Saclay--\^Ile-de-France
                   \and
                   LIX, \'Ecole polytechnique}}

\maketitle

\begin{abstract}
Model checkers use automated state exploration in order to prove
various properties such as reachability,
non-reachability, and bisimulation over state transition systems.
While model checkers have proved valuable for locating errors in
computer models and specifications, they can also be used to prove
properties that might be consumed by other computational logic
systems, such as theorem provers.
In such a situation, a prover must be able to trust that the
model checker is correct.
Instead of attempting to prove the correctness of a model checker, we
ask that it outputs its ``proof evidence'' as a formally defined
document---a proof certificate---and that this document is checked by
a trusted proof checker.
We describe a framework for defining and checking proof
certificates for a range of model checking problems.
The core of this framework is a (focused) proof system that is
augmented with premises that involve ``clerk and expert'' predicates.
This framework is designed so that soundness can be guaranteed
independently of any concerns for the correctness of the clerk and
expert specifications.
To illustrate the flexibility of this framework, we define and
formally check proof certificates for reachability and
non-reachability in graphs, as well as bisimulation and
non-bisimulation for labeled transition systems.
Finally, we describe briefly a reference checker that we have
implemented for this framework.
\end{abstract}

%%  LocalWords:  reachability bisimulation provers prover sequent

\section{Introduction}
\label{sec:introduction}

Model checkers are one way in which logic is implemented.  While one
of the strengths of model checkers is to aid in the discovery of
counterexamples and errors in specifications \cite{clarke99book}, they
can also be used to prove theorems.  Furthermore, such theorems
might be of interest to other computational logic systems such as more
general theorem provers.  One then encounters the problem of whether
or not such a theorem prover is willing to trust that model checker or
at least a particular theorem it proves.  Formally verifying a model
checker might be both extremely hard to do and undesirable especially
if that checker is being revised and improved.
A more plausible option might be to have a model checker output its
``proof evidence'' as a document (a \emph{certificate}).  If that
proof certificate can be formally checked by a trusted checker, one
might then be willing to use the theorem in a theorem prover.

Of course, model checkers are asked to solve many kinds of problems so
their proof evidence might take many different forms, ranging from
decision procedures to paths in graphs, bisimulations, traces, and
winning strategies.  If we need to have trusted checkers for all these
different kinds of proof evidence, then maybe we have not really
improved the situation of trust.  Here, we contribute to the
\emph{foundational proof certificate} (FPC) effort
\cite{miller11cpp} by providing a framework for
defining the semantics of a range of proof evidence that
naturally arises in model checking.  Such a formal semantic
model for proof evidence allows anyone to build a proof checker of
\emph{any} formally defined evidence.  Furthermore, it is possible to
have an implementation of the entire framework of FPC so that this
one system could check a wide range of proof evidence.

While this paper has a number of parallels with FPCs for first-order
logic in~\cite{chihani13cade}, that work was limited to first-order
logic \emph{without} fixed points and, as a result, that work was not
directly applicable to topics of model checking and inductive and
co-inductive theorem proving.

%%  LocalWords:  bisimulations FPC FPCs intuitionistic co-inductive
%%  LocalWords:  prover

\section{Proof theory for fixed points and certificates}
\label{sec:recent}

Having proof certificates that are foundational here means that we
need to find proof theoretic descriptions of model checking.
We shall now describe a few recent developments in proof theory that
we bring together in this paper.
%
%We shall be able to do this for simple but real model checking
%problems.
%
%Of course, since the topic of model checking is mature and varied,
%there are many of its aspects that we shall not address
%here---e.g., predicate abstraction and partial order reduction.
%
%In order to lay down a convincing and direct proof theory for model
%checking, we eschew these more advanced topic for later
%consideration.
%
Of course, the topic of model checking is mature and varied.  In order
to lay down a convincing and direct proof theory for model checking,
we eschew many of its more advanced topics---e.g., predicate
abstraction and partial order reduction---for later consideration.

\subsection{Fixed points as defined predicates}

One of the earliest applications of sequent calculus to computational
logic was to provide an execution model for logic programming
\cite{miller91apal}.  That analysis, however, supported only the
``open world assumption'' of logic programming:
negation-as-finite-failure was not touched by that work.
Schroeder-Heister \cite{schroeder-heister93lics}
and Girard \cite{girard92mail} showed how sequent calculus could be
extended with inference rules for fixed points (or \emph{defined}
predicates), thereby embracing important aspects of the \emph{closed
world assumption} and negation-as-finite-failure.  The key additions
to sequent calculus were rules for unfolding fixed point expressions
as well as dealing with equality over the Herbrand universe.
A series of papers \cite{gacek11ic,mcdowell00tcs,miller05tocl} added
induction and co-induction to the sequent calculi for intuitionistic
and classical logics.  Those papers have been used to design the
Bedwyr model checker \cite{baelde07cade,tiu05eshol} and the Abella
interactive theorem prover \cite{baelde14jfr}.

Fixed point expressions will be written as $\mu{}B\,\t$ or
$\nu{}B\,\t$, where $B$ is an expression representing a higher-order
abstraction, and $\t$ is a list of terms.  The unfolding of the fixed
point expression $\mu B\,\t$ is written as
$B(\mu{}B)\,\t$.  It is important to understand that we shall treat
both $\mu$ (least fixed point operator) and $\nu$ (greatest fixed
point operator) as logical connectives since they will have
introduction rules: they are also de Morgan duals of each other.

% XXX clear?  DM Not so clear to me.
% Throughout this paper, we refer to fixed point expressions as
% \emph{defined atoms}, written as ...
% The constants $\mu$ and $\nu$ are still considered as connectives, in the sense that
% the equality of the defined atoms $\mu{}B_1\,\t_1$ and
% $\mu{}B_2\,\t_2$ is the syntactic equality of the pairs $(B_1,\t_1)$
% and $(B_2,\t_2)$.

%%  LocalWords:  Heister co-induction intuitionistic Herbrand prover
%%  LocalWords:  duals

\subsection{Fixed points in linear logic}

Surprisingly, it is linear logic and not intuitionistic or classical
logics per se that is most relevant to our exposition on model
checking.  The logic MALL (\emph{multiplicative additive linear logic})
is an elegant, small logic that is, in and of itself, not appropriate
for formalizing mathematics and computer science since it is not
capable of modeling unbounded behaviors (for example, it is
decidable).  While Girard extended MALL with the ``exponentials'' ($!$ and
$?$) \cite{girard87tcs},  Baelde \cite{baelde12tocl}
extended it by adding the least ($\mu$) and greatest
($\nu$) fixed points operators as logical connectives.  The resulting
logic, called \mumall, forms the proof theoretic foundation of this
paper.

To make the use of linear logic easier to swallow for those more
familiar with model checking, we adopt the following shallow changes
to its presentation.
First, we use a two-sided sequent calculus instead of the one-sided
calculus used for \mumall.
While this change will double the size of our proof system, it will
make inference rules look more familiar.
Second, we replace the linear logic connectives with familiar
connectives (although with annotations).
In particular, we replace $\lltens$, $\llwith$, $\llplus$ and their
units $\llone$, $\top$ and $\llzero$ with $\pand$, $\nand$, $\por$,
$\ptrue$, $\ntrue$ and $\pfalse$, respectively.
(Truth functionally, the two versions of these operators are
equivalent: their differences only influence the structure of focused
proofs.)
We also replace the negatively biased false $\bot$ with $\nfalse$, and
instead of the multiplicative disjunction $A\llpar B$, we use the
implication $A^\perp\nimp B$: the de Morgan dual of $A\nimp B$ is
$A\pand B^\perp$.   Negation is written as $\cdot\nimp\nfalse$.

In addition, we consider $\mu$ as positive and $\nu$ as negative; this
arbitrary choice has been shown to give a convenient natural
interpretation to the structure of focused proofs \cite{baelde12tocl}.
We therefore have the negative connectives
$\nfalse$, $\nimp$, $\ntrue$, $\nand$, $\forall$, $\neq$ and $\nu$,
and the positive connectives
$\ptrue$, $\pand$, $\pfalse$, $\por$, $\exists$, $=$ and $\mu$.

%%  LocalWords:  se intuitionistic Girard exponentials

\subsection{Focused proof systems}
\label{sec:focused}

In order to have the kind of control we need to support a definable
notion of proof certificate, we make use of a \emph{focused proof
  system}.  Such sequent calculus proof systems are built from
alternating phases which allow us to define flexible proof building
protocols that can be used to drive proof search.  During the
\emph{asynchronous} phase of proof building, simple (invertible)
computations build a proof and during the \emph{synchronous} phase,
information needed for the construction of a proof (such as which
branch of a disjunction to prove) must be found.

Focusing requires polarizing all formulas as being either negative or
positive.  A formula is negative or positive according to its
top-level connective, and it is \emph{purely negative} (resp.
\emph{purely positive}) when its connectives are positive if, and only
if, they occur under an odd (resp. even) number of implications.
Notice that the de Morgan dual of a positive (resp. purely positive)
formula is a negative (resp. purely negative) formula.
We call a formula \emph{bipolar} when it is made of purely negative
(resp. positive) subformulas occurring under an even (resp. odd)
number of implications in a purely negative context.

Focusing also relies on the sequents having additional storage zones
on each side of the turnstile, where formulas can be stored and left
untouched by logical inference rules.  For instance, the usual
one-sided focused presentation of \mumall{} \cite{baelde12tocl}
has one of these zones, similarly
to the focused proof system for linear logic given by Andreoli
\cite{andreoli92jlc}.
A two-sided subsystem of \mumallf{}, called \muF{},
makes use of two storage zones, noted $\Nscr$ and $\Pscr$,
which are lists of, respectively, negative and positive formulas.
(\Cref{sec:example} contains an example of a \muF{} proof.)
%
%
%We consider \muF, a two-sided subsystem of
%\mumallf{} making use of two storage zones, one on each side of the
%turnstile.
%The outer storage zones are noted $\Theta$ and $\Upsilon$ and are
%referred to as \emph{tables}; they are sets of %defined atoms.
%pairs $(i,D)$, where $D$ is a defined atom and $i$ is a hash of some
%sorts that we call \emph{index}.
%The median storage zones are noted $\Nscr$ and $\Pscr$ and assume the
%These are noted $\Nscr$ and $\Pscr$ and assume the
%role of the storage zone in \mumallf;
Between the arrows and the turnstile, are the contexts
$\Gamma$ and $\Delta$: these
are lists of formulas in (unfocused) $\Uparrow$-sequents, and
sets of up to one formula in (focused) $\Downarrow$-sequents.
The sequents of the \muF{} system are therefore:
\begin{align*}
  {\Nscr}\Uparrow{\Gamma}\vdash{} & {\Delta}\Uparrow{\Pscr} & &
    \text{unfocused, similar to the \mumallf{} sequent }
    \vdash\Nscr^\perp,\Pscr\Uparrow\Gamma^\perp,\Delta \\
  \Downarrow{A}\vdash{} & & &
    \text{left-focused, similar to }
    \vdash\Downarrow{}A^\perp \\
  {}\vdash{} & A\Downarrow{} & &
    \text{right-focused, similar to }
    \vdash\Downarrow{}A
\end{align*}
%\certcolour{}
%\[\begin{array}{cl}
%  \async{\Theta}{\Nscr}{\Gamma}{\Delta}{\Pscr}{\Upsilon} &
%    \text{unfocused, similar to the \mumallf{} sequent }
%    \vdash\Nscr^\perp,\Pscr\Uparrow\Gamma^\perp,\Delta \\
%  \syncL{\Theta}{A}{\Upsilon} &
%    \text{left-focused, similar to }
%    \vdash\Downarrow{}A^\perp \\
%  \syncR{\Theta}{A}{\Upsilon} &
%    \text{right-focused, similar to }
%    \vdash\Downarrow{}A
%\end{array}\]
%\certcolour{thecertcolour}

%%  LocalWords:  subformulas invertible Andreoli

\subsection{Foundational proof certificates}
\label{sec:fpc}

If we think of the implementers of computational logic systems (\eg,
model checking systems) as our clients, our job in this project is
to formally check our client's proof evidence for formal correctness.
Our approach is to have this evidence translated into a sequent
calculus proof.
% of a desired property; once we can get our checker to
% go over the proof, we have a very high degree of confidence in it.
Of course, we would not dream of asking our clients to supply a
sequent calculus proof in the first place: such proofs are often huge,
too messy, and too esoteric.
Instead, we want to take from our clients objects with which they are
familiar (\eg, paths, simulations, \etc) and find flexible and
high-level means to have our framework extract
information from those objects in order to trace out a complete
formal sequent calculus proof.

To this intent, \Cref{fig:muFa0,fig:muFa1} present \muFa{}, which is a
version of \muF{} augmented with a term $\Xi$ (encoding an actual
certificate) as well as with \emph{clerk} and \emph{expert}
predicates (examples of which we provide soon).
This augmentation has two components.
First, every sequent (either $\Uparrow$
or $\Downarrow$) is given an extra argument we write as $\cert{\Xi}$.
Thus, sequents now display as
\begin{align*}
  & \qquad & \qquad
  & \async[\Xi]{\Theta}{\Nscr}{\Gamma}{\Delta}{\Pscr}{\Upsilon}, &
  & \syncL[\Xi]{\Theta}{A}{\Upsilon}, &
  & \text{and} &
  & \syncR[\Xi]{\Theta}{A}{\Upsilon}.
  & \qquad & \qquad
\end{align*}
Second, every inference rule is given an
additional premise.  In all cases, this premise is an atomic formula
with either a clerk or expert predicate as its head
symbol: if the conclusion of the inference rule is a
$\Downarrow$-sequent, then the premise atom uses an expert predicate
(noted $\cert{\star_e(\dots)}$ for the rule $\star$); otherwise, the
conclusion is an $\Uparrow$-sequent and the atom uses a clerk
predicate (noted $\cert{\star_c(\dots)}$).

In the case of the clerk rules, the
premise atom relates the $\Xi$ value of the concluding sequent with
the corresponding value of $\Xi$ for all premises: \eg,
\[ \porl \]
In this way, the certificate $\Xi$, intended to aid in the proof of
the concluding sequent, can be transformed into two certificates that
are used to prove the two premise sequents.  We refer to the
predicates used in the asynchronous phase as clerks since these
predicates do not need, in general, to examine the actual information
%in the proof certificate (except for the induction, co-induction and cut
in the proof certificate (except for the induction and co-induction
rules, there is no consumption of information during the asynchronous
phase).  Instead, the clerks are responsible for keeping track of how
a proof is unfolding: for example, $\Xi_1$ might be a copy of $\Xi_0$
but with the fact that checking has moved to the left premise instead
of the right premise.

Experts are responsible for extracting information from a certificate.
For example, \muFa{} contains the inference rule
\[
  \existsr
\]
Notice here that the exists-expert $\existse(\cdot,\cdot,\cdot)$ not
only computes the continuation certificate $\Xi_1$ but also a term $t$
to be used to witness the existential variable.

The exact nature of both certificate terms $\Xi$ and of the clerk and
expert predicates is not important to guarantee soundness of this
system.  That is, no matter how certificates, clerks, and experts are
specified, if there is a proof in \muFa{} then there is a proof in
\muF{} of the same sequent, which can be obtained by deleting from the
proof in \muFa{} all references to $\Xi$, including the additional
premises.  Notice also that experts are not required to act
particularly expertly: it is entirely possible for the
$\existse(\Xi_0,\Xi_1,t)$ premise to functionally determine one $t$
from $\Xi_0$, or to relate all terms $t$ to $\Xi$.  In the latter case,
the actual value of $t$ used in a successful \muFa{} proof is
determined from other aspects of the proof checking process (typically
implemented using unification).

%%  LocalWords:  co-induction FPC intuitionistic Frege Herbrand

\section{A proof system underlying model checking}
\label{sec:proof system}

FPCs were first proposed in~\cite{chihani13cade,miller11cpp} in the
context of first-order logic and were used successfully to define and
check proof evidence in the form of resolution refutations, Herbrand
instances (expansion trees), natural deduction ($\lambda$-terms),
Frege proofs, \etc{}
We shall now adapt this approach to formally define the semantics of a
range of proof evidence that can arise in simple but real model
checking problems.

We shall later illustrate just how such a formal semantics can be
provided for the following four kinds of proof evidence.
These particular examples have been selected for their universality:
numerous problems in model checking are related to them.
\begin{enumerate}
  \item The fact that two nodes are related by the transitive closure
    of a graph's adjacency relation can be witnessed by an
    \emph{explicit path} through the graph.

  \item The fact that two nodes are \emph{not} related by transitivity
    can be witnessed by pointing out that the \emph{reachable set} of
    one does not contain the other.

  \item Given an LTS (labeled transition system), the fact that two
    nodes are similar/bisimilar can be witnessed by a set of pairs
    called \emph{simulation}/\emph{bisimulation}.

  \item If two nodes in an LTS are \emph{not} bisimilar, then there is
    a \emph{Hennessy-Milner logic (HML) formula} that is satisfied by
    one but not by the other.
\end{enumerate}

\begin{figure}
  \certcolour{thecertcolour}
  \textsc{Asynchronous connective introductions}
  \[ \eqsl \qquad \neqfr \]
  \[ \ptruel \qquad \nfalser \]
  \[ \pandl \qquad \nimpr \]
  \[ \eqfl \qquad \neqsr \]
  \[ \pfalsel \qquad \ntruer \]
  \[ \porl \]
  \[ \nandr \]
  \[ \existsl \qquad \forallr \]

  \textsc{Synchronous connective introductions}
  \[ \neqfl \qquad \eqsr \qquad \nfalsel \qquad \ptruer \]
  \[ \nimpl \qquad \pandr \]
  \[ \nandl \qquad \porr \]
  \[ \foralll \qquad \existsr \]

  \textsc{Structural rules}
  \[ \storel \qquad \storer \]
  \[ \decidel \qquad \decider \]
  \[ \releasel \qquad \releaser \]

  \caption{The $\muFa_0$ proof system. (This proof system is best viewed using color).\newline
    $y$ stands for a fresh eigenvariable, $s$ and $t$ for terms, $N$
    for a negative formula, $P$ for a positive formula, and $C$ for
    the abstraction of a formula over a variable.\newline
    The $\dag$ proviso requires that $\theta$ is the \emph{mgu} of
    $s$ and $t$, and the $\ddag$ proviso requires that $s$ and $t$
    are not unifiable.
  }
  \label{fig:muFa0}
\end{figure}

\begin{figure}
  \certcolour{thecertcolour}
  \textsc{Fixed-point rules}
  \[ \ind \]
  \[ \coind \]
  \[ \mul \qquad \nur \]
  \[ \nul \qquad \mur \]

  %\textsc{Interface rules}
  %\[ \cutl \]
  %\[ \cutr \]
  %\[ \initl \qquad \initr \]

  \caption{The \muFa{} proof system results from adding these rules to
    $\muFa_0$.\newline
    $\y$ stands for an list of fresh eigenvariables, $\t$ for an
    list of terms, and $B$ for the abstraction of a formula over a
    predicate and a variable list.
    %, $D$ for a defined atom
  }
  \label{fig:muFa1}
\end{figure}

\subsection{Core proof system}
\label{sec:core}

\Cref{fig:muFa0,fig:muFa1} contain the rules for the augmented focused
proof systems $\muFa_0$ and \muFa{}.  One could obtain the
non-augmented systems $\muF_0$ and \muF{} by ignoring the certificates
(annotated $\Xi$ variables) and the clerk and expert premises;
the resulting rules would be no more than (slightly
restricted) two-sided versions of the \mumallf{} rules.
%The first four groups of rules
%are mainly two-sided versions of the rules from \mumallf.
%The last group provides the only ``context dependent'' means of
%completing a proof.
The various clerk and expert predicates
are named and displayed in their corresponding inference rules.
Notice that those inference rules that involve the use of
eigenvariables (${\exists_L}$, ${\forall_R}$, ${\mu}$ and ${\nu}$)
require the associated clerk predicates to return abstractions over
certificates: in this way, premise certificates can be applied to the
eigenvariables.

A key element of our proof theoretic treatment of model checking via
\muFa{} is the fact that %, tables notwithstanding,
focused sequents
contain only one formula.  This fact entails that \muFa{} can only be
complete with respect to \mumallf{} on a fragment where derivations
satisfy this constraint.  In particular, the $\Nscr$ and $\Pscr$ zones
must never contain more than one formula, and never both at the same
time.  This can be ensured at least for the $\muFa_0$ subsystem by the
following restriction on formulas.

\begin{definition}[\switchable{} formula, \switchable{} occurrence]
  A \muFa{} formula is \emph{\switchable{}} if
  %\vspace{-\topsep}
  \begin{itemize}
    \item whenever a subformula $C\pand{}D$ occurs negatively (under
      an odd number of implications), either $C$ or $D$ is purely
      positive;
    \item whenever a subformula $C\nimp{}D$ occurs positively (under
      an even number of implications), either $C$ is purely positive or
      $D$ is purely negative.
  \end{itemize}
  %\vspace{-\topsep}
  An occurrence of a formula $B$ is \emph{\switchable{}} if it appears on
  the right-hand side (resp. left-hand side) and $B$ (resp.
  $B\nimp\nfalse$) is \switchable{}.
\end{definition}
Notice that both a purely positive formula and its de Morgan dual are
\switchable{}.  The follow theorem is proved by a simple induction on
the structure of $\muFa_0$ proofs.

\begin{theorem}[switchability]
  \label{thm:switch}
  Let $\Pi$ be a $\muFa_0$ derivation of either
  $\async{}{}{A}{}{}{}$ or $\async{}{}{}{A}{}{}$, where the occurrence
  of $A$ is \switchable{}.
  Every sequent in $\Pi$ that is the conclusion of a rule that
  switches phases (either a decide or a release rule) contains exactly
  one occurrence of a formula and that occurrence is \switchable{}.
\end{theorem}

\Cref{thm:switch} states that an invariant of the $\muFa_0$ proof system
(for \switchable{} theorems) is that the number of non-purely
asynchronous formulas (\ie non-purely positive from $\Nscr$ and
$\Gamma$, and non-purely negative from $\Pscr$ and $\Delta$) is one or
less.
Keeping sequents mostly asynchronous allows the
asynchronous phase to deal with most of the context: that way the
synchronous phase is left with a single, meaningful formula.
(The structure of focused proofs based on switchable formulas is
similar to the structure of \emph{simple games} in the game-theoretic
analysis of focused proofs in~\cite[Section 4]{delande10apal}.)
While the restriction to switchable formulas provides a match to the
model checking problems we develop here, that restriction is not
needed for using clerks and experts (the examples in
\cite{chihani13cade} involve non-switchable formulas).

%%  LocalWords:  subformula muFa eigenvariables thm

\subsection{Encoding of recursively defined predicates}

In order to exploit the properties of $\muFa_0$ in model checking
problems, we need them to extend to \muFa{} by adding
%problems, we need them to extend to \muFa, especially to the
%fixed-point and cut rules.
fixed-point rules.  As those rules make use of the higher-order
variables $S$ (an invariant which is either a pre-fixed point or
a post-fixed point) and $B$ (the body of a predicate definition), they
cannot be used freely without violating \Cref{thm:switch}.
We propose the following constraints on \muFa{} proofs of \switchable{}
formulas so as to have exactly one formula per sequent when phases
are switched:
\begin{itemize}
  \item ``arithmetic'' restriction: $S$ and $B$ are purely positive
    (resp.\ negative);
  \item ``model checking'' restriction: $S$ is purely negative (resp.\
    positive), and the context does not trigger synchronous rules
    ($\Nscr$ is empty, $\Gamma$ is purely positive and $\Delta$ is
    purely negative).
\end{itemize}
The former restriction would allow to extend the scope of the
framework by handling simple theorems involving inductive definitions
(\eg{} about natural numbers), but is not treated here.
% XXX "better suits our needs"?
The latter restriction better suits our needs (since an asynchronous
context fits the spirit of model checking) and is respected by all our
examples.

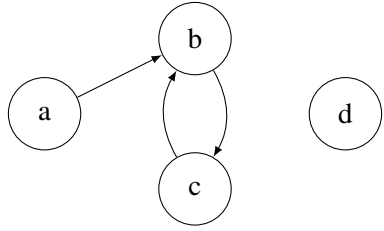
\begin{figure}[t]
  \hspace{\stretch{1}}
  \begin{tikzpicture}[%
      every path/.style={>=latex},%
    ]
    \node[state] (a) at (0,0)  { a };
    \node[state] (b) at (2,1)  { b };
    \node[state] (c) at (2,-1) { c };
    \node[state] (d) at (4,0)  { d };

    \draw[->] (a) edge (b);
    \draw[->] (b) edge[bend left] (c);
    \draw[->] (c) edge[bend left] (b);
  \end{tikzpicture}
  \hspace{\stretch{1}}
  \caption{(Un)reachability problem}
  \label{fig:graph}
\end{figure}

\begin{example}
  \label{ex:graph}
  Horn clauses (in the sense of Prolog) can be encoded as purely
  positive fixed point expressions.  For example, here is the Horn
  clause logic program (using the $\lambda$Prolog syntax, the
  \verb+sigma Y\+ construction encodes the quantifier $\exists{}Y$)
  for specifying the graph in \Cref{fig:graph} and its transitive
  closure:
\begin{verbatim}
step a b.   step b c.    step c b.
path X Y :- step X Y.    path X Z :- sigma Y\ step X Y, path Y Z.
\end{verbatim}
  We can translate the \verb.step. relation into the binary predicate
  $\one{\cdot}{}{\cdot}$ defined by
  \begin{align*}
    \mu\left(\lambda{}A\lambda{}x\lambda{}y.\,
    ((x=a)\pand(y=b))\por((x=b)\pand(y=c))\por((x=c)\pand(y=b))\right)
  \end{align*}
  which only uses positive connectives.  Likewise, \verb.path. can be
  encoded as $\ppathv$:
  \begin{align*}
    \mu\left(\lambda{}A\lambda{}x\lambda{}z.\,
    \one{x}{}{z}\por(\exists{}y.\,\one{x}{}{y}\pand{}A\,y\,z)\right)
  \end{align*}
\end{example}

In general, it is sensible to view any purely positive least fixed
point expression as a predicate specified by Horn clauses.
(For example, SOS rules for CCS are easily seen as Horn clauses.)

\begin{example}
  \label{ex:lts}
  Let the ternary predicate $\one{\cdot}{\cdot}{\cdot}$ describe a
  labeled transition system.  It can be defined as a purely positive
  fixed point expression of the form
  \begin{align*}
    \mu\left(\lambda{}A\lambda{}p\lambda{}a\lambda{}q.\,\bigpor\strut_i((p=u_i)\pand(a=v_i)\pand(q=w_i))\right)
  \end{align*}
  and the simulation and bisimulation relations can be defined
  as the following greatest fixed point expressions (note: the second
  contains both $\nand$ and $\pand$).  Both of these formulas are switchable.
  \begin{align}
    \nu\big(\lambda{}S\lambda{}p\lambda{}q.\, &
      \forall{}a\forall{}p'.\,\one{p}{a}{p'}\nimp
      \exists{}q'.\,\one{q}{a}{q'}\pand{}S\,p'\,q'
      \big)\tag{$\ssimv$} \\
    \begin{split}
      \nu\big(\lambda{}B\lambda{}p\lambda{}q.\, &
        (\forall{}a\forall{}p'.\,\one{p}{a}{p'}\nimp
        \exists{}q'.\,\one{q}{a}{q'}\pand{}B\,p'\,q') \\
      \nand &
        (\forall{}a\forall{}q'.\,\one{q}{a}{q'}\nimp
        \exists{}p'.\,\one{p}{a}{p'}\pand{}B\,q'\,p')\big)
    \end{split}
    \tag{$\bisimv$}
  \end{align}
%
%  Notice that the latter contains occurrences of both $\nand$ and
%  $\pand$.
\end{example}

%%  LocalWords:  co-induction pre Prolog CCS bisimulation

\subsection{Common proof certificates}
\label{sec:common}

The presentation of an FPC now involves the following three steps.
\begin{enumerate}
  \item Describe how unpolarized formulas should be polarized.
  \item Describe the structure of certificates $\Xi$.  This can be
    done, for example, by describing the signature of constructors for
    certificates.
  \item Define the clerk and expert predicates.
\end{enumerate}
To ease steps $2$ and $3$, we define the following certificate
constructors (shown together with their types), which describe generic
focused proof behaviors.
The associated clauses can be included into any subsequent clerks and
experts definitions.

The \textbf{\stopcert\texttt{:cert}}
certificate authorizes no search; it is to be used as a
continuation certificate for other certificate constructors.

The \textbf{\synccert\texttt{:cert->cert}}
certificate constructor authorizes \muFa{} to conduct an
unbounded synchronous search for a proof before handing the search
over to a continuation certificate.  It has no clerks and its experts
run an exhaustive non-deterministic search for $\por$ and $\exists$.
The experts for the right rules are:
\begin{align*}
               & \eqse(\synccert(\Xi)). &
               & \pore(\synccert(\Xi),\synccert(\Xi),1). \\
               & \pande(\synccert(\Xi),\synccert(\Xi),\synccert(\Xi)). &
               & \pore(\synccert(\Xi),\synccert(\Xi),2). \\
  \forall{}T.\,& \existse(\synccert(\Xi),\synccert(\Xi),T). &
               & \mure(\synccert(\Xi),\synccert(\Xi)). \\
               & \releasere(\synccert(\Xi),\Xi).
\end{align*}

The \textbf{\asynccert\texttt{:cert->cert}}
certificate constructor is the dual of \synccert; it handles an
asynchronous phase and has no experts apart from the decide rules.
The clerks for the left rules are:
\begin{align*}
  & \eqsc(\asynccert(\Xi),\asynccert(\Xi)). &
  & \porc(\asynccert(\Xi),\asynccert(\Xi),\asynccert(\Xi)). \\
  & \pandc(\asynccert(\Xi),\asynccert(\Xi)). \\
  & \existsc(\asynccert(\Xi),\lambda{}x.\,\asynccert(\Xi)). &
  & \mule(\asynccert(\Xi),\asynccert(\Xi)). \\
  & \storelc(\asynccert(\Xi),\asynccert(\Xi)). &
  & \decidele(\asynccert(\Xi),\Xi).
\end{align*}

\textbf{\bipolecert[n]\texttt{:cert}}
is actually short-hand for a chain of
$n$ $\asynccert{(\synccert{(\cdot)})}$ before a final \stopcert.
%$n$ alternating \asynccert{} and \synccert{} before a final \stopcert.
It is used for bounded-depth search when simple search strategies
would otherwise not terminate.  We also write \textbf{\bipolecert\texttt{:cert}} for
$\bipolecert_1=\asynccert(\synccert(\stopcert))$.

The \textbf{\deccert\texttt{:cert}}
constructor is short-hand for \bipolecert[\infty], the unbounded
version of \bipolecert[n].
It is a general purpose decision procedure used for automated and
unguided proving.
Its rules are similar to those from \synccert{} and \asynccert{}, and
can be obtained via the equivalence
$\deccert=\asynccert{(\synccert{(\deccert)})}$.

The two constructors \textbf{\invcert}
and \textbf{\coinvcert\texttt{:(i->i->bool)->cert->cert}} each
take an explicit
predicate $\Sscr$ as parameter.  It is expected to be proved to be an
invariant with the help of \bipolecert.
\begin{align*}
  & \forall\Sscr.\,\inde(\invcert(\Sscr,\Xi),\lambda\x.\,\bipolecert,\Xi,\Sscr) &
  & \forall\Sscr.\,\coinde(\coinvcert(\Sscr,\Xi),\Xi,\lambda\x.\,\bipolecert,\Sscr)
\end{align*}

We now turn our attention to describing how to formally define the
four kinds of proof evidence mentioned earlier in \Cref{sec:proof
system}.  Some of the constructors defined above will be used in those
definitions.

%%  LocalWords:  bisimulation Milner bisimulations reachability HML

\section{Examples: certificates for graphs}

We use the notations from \Cref{ex:graph} to define
$\one{\cdot}{}{\cdot}$ and $\ppathv$.

\subsection{Lists as reachability certificates}
\label{sec:reach}

The natural choice for a certificate of the proof of
$\vdash\ppath{x}{y}$ is an explicit path, \ie a list of nodes starting
right after $x$ and ending right before $y$.   In fact, this list $L$
can be used directly as the proof certificate.
Aside from the initial %$\forall{}L.\,\storerc(L,L).$
$\storerc$, no clerks are invoked in the process of checking this
particular FPC.
The following clauses defining the experts only use the provided
information to instantiate the logical variables of the proof.
\begin{align*}
  \forall{}X\forall{}L.\,& \existse(X::L,L,X). &
  \forall{}L.\,          & \pande(L,\synccert(\stopcert),L).  &
  \forall{}L.\,          & \decidere(L,L). \\
  \forall{}X\forall{}L.\,& \pore(X::L,X::L,2). &
  \forall{}L.\,          & \pore(nil,\synccert(\stopcert),1). &
  \forall{}L.\,          & \mure(L,L).
\end{align*}
In this setting, the $\synccert(\stopcert)$ certificate will terminate
quickly since it is only searching through the term that defines
$\one{\cdot}{}{\cdot}$.
\begin{example}
  In \Cref{fig:graph}, $(c)$ is reachable from $(a)$, as
  witnessed by certificates like $[b]$, $[b;c;b]$, \etc
\end{example}

%%  LocalWords:  Reachability Un reachability Prolog FPC

\subsection{Invariants as non-reachability certificates}
\label{sec:nonreach}

The non-reachability problem comes in two forms: if there are no loops
in the graph, then a simple check of the set of nodes reachable from
the first node provides a simple decision procedure; if there are
loops, then induction is needed.

In the first case, the decision procedure can directly be translated
as an FPC for proving $\vdash\neg\ppath{x}{y}$.
\begin{example}
  In \Cref{fig:graph}, $(a)$ is not reachable from $(d)$,
  as witnessed by $\asynccert(\stopcert)$.
\end{example}

On the other hand, if the underlying graph has loops, then the rules
of \Cref{fig:muFa0} only do not allow proof search to terminate.
As the body $B$ of the $\ppathv$ expression (\ie, the displayed
formula without $\mu$) is purely positive, a bipole can prove that a
chosen purely negative predicate $\Sscr$ containing no fixed point
expressions is an induction invariant
($\certcolour{thecertcolour}\async[\bipolecert]{}{}{B\,\Sscr\,x\,y}{\Sscr\,x\,y}{}{}$),
which means that we can use the certificate constructor
$\invcert(\Sscr,\cdot)$.
Then we use another bipole as continuation certificate for this
constructor to check that the invariant is adequate for the
refutation of $\ppath{t}{u}$
($\certcolour{thecertcolour}\async[\bipolecert]{}{}{S\,t\,u}{\cdot}{}{}$).

Here, the invariant can be chosen so as to represent the fact of
\emph{not} belonging to the set $\mathcal{T}\times\{u\}$, where
$\mathcal{T}$ is the reachable set of $\{t\}$.

\begin{example}
  In \Cref{fig:graph}, $(d)$ is not reachable from $(b)$, as witnessed by
  $\invcert(\Sscr,\bipolecert)$, where the invariant $\Sscr$ (built
  from the set $\{b,c\}\times\{d\}$) is
  \[\lambda{}x\lambda{}y.\,((x=b\pand{}y=d)\por(x=c\pand{}y=d))\nimp\nfalse\]
\end{example}

%%  LocalWords:  reachability FPC asyncL bipole invcert

\section{Examples: certificates for labeled transition systems}

Bisimilarity and similarity are important relationships in the domain
of process calculus and model checking.  To illustrate how these can
be captured as FPCs, we first restrict our attention to the existence
of a simulation between finite labeled transition systems;
bisimilarity is then addressed by expanding on this presentation.
We define $\one{\cdot}{\cdot}{\cdot}$ (for the LTS), $\ssimv$ (for
simulated-by) and $\bisimv$ (for bisimulated-by) as seen in
\Cref{ex:lts}.

\subsection{Invariants as simulation certificates}
\label{sec:sim}

We shall consider two cases: one where the underlying transition
system is noetherian and one where it is not.  An LTS is said to be
\emph{noetherian} if there is no infinite sequence of transitions
$\one{p_1}{a_1}{\one{p_2}{a_2}{\cdots}}$ (in the setting of finite
LTSs, this is equivalent to the absence of loops).

In the noetherian case, there is a decision procedure to determine
whether or not one process is simulated by another: one simply
attempts to incrementally check simulation at every point.  This
systematic search can be described using the clerks and experts of the
\deccert{} certificate, which allows a proof to be built from any
number of bipoles (one for each unfolding of the simulation predicate,
which formula is itself bipolar).

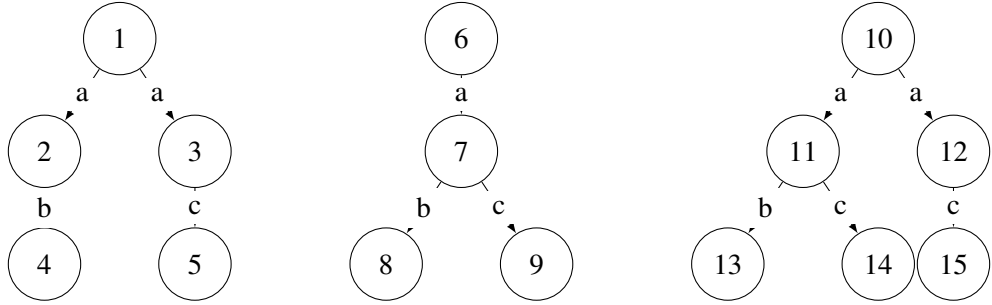
\begin{figure}[t]
  \hspace{\stretch{1}}
  \begin{tikzpicture}[%
      every path/.style={>=latex},%
      lbl/.style={fill=white},%
    ]
    \node[state] (x1) at (0,0)     { 1 };
    \node[state] (x2) at (-1,-1.5) { 2 };
    \node[state] (x3) at (1,-1.5)  { 3 };
    \node[state] (x4) at (-1,-3)   { 4 };
    \node[state] (x5) at (1,-3)    { 5 };

    \draw[->] (x1) edge node[lbl] { a } (x2);
    \draw[->] (x1) edge node[lbl] { a } (x3);
    \draw[->] (x2) edge node[lbl] { b } (x4);
    \draw[->] (x3) edge node[lbl] { c } (x5);
  \end{tikzpicture}
  \hspace{\stretch{1}}
  \begin{tikzpicture}[%
      every path/.style={>=latex},%
      lbl/.style={fill=white},%
    ]
    \node[state] (x6) at (0,0)    { 6 };
    \node[state] (x7) at (0,-1.5) { 7 };
    \node[state] (x8) at (-1,-3)  { 8 };
    \node[state] (x9) at (1,-3)   { 9 };

    \draw[->] (x6) edge node[lbl] { a } (x7);
    \draw[->] (x7) edge node[lbl] { b } (x8);
    \draw[->] (x7) edge node[lbl] { c } (x9);
  \end{tikzpicture}
  \hspace{\stretch{1}}
  \begin{tikzpicture}[%
      every path/.style={>=latex},%
      lbl/.style={fill=white},%
    ]
    \node[state] (x10) at (0,0)     { 10 };
    \node[state] (x11) at (-1,-1.5) { 11 };
    \node[state] (x12) at (1,-1.5)  { 12 };
    \node[state] (x13) at (-2,-3)   { 13 };
    \node[state] (x14) at (0,-3)    { 14 };
    \node[state] (x15) at (1,-3)    { 15 };

    \draw[->] (x10) edge node[lbl] { a } (x11);
    \draw[->] (x10) edge node[lbl] { a } (x12);
    \draw[->] (x11) edge node[lbl] { b } (x13);
    \draw[->] (x11) edge node[lbl] { c } (x14);
    \draw[->] (x12) edge node[lbl] { c } (x15);
  \end{tikzpicture}
  \hspace{\stretch{1}}
  \caption{Non-(bi)similar noetherian labeled transition systems}
  \label{fig:LTS1}
\end{figure}

\begin{example}
  In \Cref{fig:LTS1}, the process $(1)$ is simulated by the process $(6)$, as
  witnessed by the certificate \deccert.
\end{example}

In the more general (possibly non-noetherian) setting, we need to recall
the formal definition of the simulation relation as a set.  A binary
relation $\Sscr$ is a simulation if whenever $\tup{p,q}\in\Sscr$ and
whenever $\one{p}{a}{p'}$ holds, then there exists a $q'$ such that
$\one{q}{a}{q'}$ holds and $\tup{p',q'}\in\Sscr$.  We say that process
$p$ is simulated by process $q$ if there is a simulation $\Sscr$ such
that $\tup{p,q}\in\Sscr$.

Let $\Sscr$ be a finite set of pairs and let $\hat\Sscr$ be the purely
positive expression
$\lambda{}x\lambda{}y.\,\bigpor_{\tup{p,q}\in\Sscr}(x=p\pand y=q)$.
As the body $B$ of the $\ssimv$ expression is a bipolar formula, a
bipole can prove the closure condition for (finite) simulations
($\certcolour{thecertcolour}\async[\bipolecert]{}{}{\hat\Sscr\,x\,y}{B\,\hat\Sscr\,x\,y}{}{}$),
so we can use the certificate constructor
$\coinvcert(\hat\Sscr,\cdot)$.  Once again, we use another bipole
as continuation certificate to complete the proof that $p$ is
simulated by $q$
($\certcolour{thecertcolour}\async[\bipolecert]{}{}{\cdot}{\hat\Sscr\,p\,q}{}{}$).

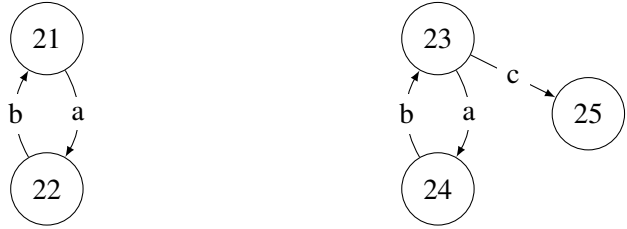
\begin{figure}[t]
  \hspace{\stretch{1}}
  \begin{tikzpicture}[%
      every path/.style={>=latex},%
      lbl/.style={fill=white},%
    ]
    \node[state] (x21) at (0,1)  { 21 };
    \node[state] (x22) at (0,-1) { 22 };

    \draw[->] (x21) edge[bend left] node[lbl] { a } (x22);
    \draw[->] (x22) edge[bend left] node[lbl] { b } (x21);
  \end{tikzpicture}
  \hspace{\stretch{1}}
  \begin{tikzpicture}[%
      every path/.style={>=latex},%
      lbl/.style={fill=white},%
    ]
    \node[state] (x23) at (0,1)  { 23 };
    \node[state] (x24) at (0,-1) { 24 };
    \node[state] (x25) at (2,0)  { 25 };

    \draw[->] (x23) edge[bend left] node[lbl] { a } (x24);
    \draw[->] (x24) edge[bend left] node[lbl] { b } (x23);
    \draw[->] (x23) edge            node[lbl] { c } (x25);
  \end{tikzpicture}
  \hspace{\stretch{1}}
  \caption{Non-noetherian labeled transition systems}
  \label{fig:LTS2}
\end{figure}

\begin{example}
  According to \Cref{fig:LTS2}, the set $\{(21,23),(22,24)\}$ is a
  simulation and, therefore, the process $(21)$ is simulated by the
  process $(23)$.  This corresponds to the following certificate
  \[\coinvcert\left(\lambda{}x\lambda{}y.\,(x=21\pand{}y=23)\por(x=22\pand{}y=24),\bipolecert\right).\]
\end{example}

Providing an entire invariant as part of a proof certificate or
restricting to the case when an invariant is finite certainly limits
what kinds of simulation relationships can be proved.  In general,
invariants will not be finite and, even when they are, they are
large.  It is for reasons such as this that there has been a
great deal of work on bisimulation-up-to
\cite{milner89book,pous11atbc}: generally, it is possible to discover
and check a closure property of a much smaller relationship and then
via various meta-theoretic properties, ensure that such closure
properties entail the existence of a proper (bi)simulation.

%%  LocalWords:  Bisimilarity FPCs sim noetherian LTS FPC
%%  LocalWords:  co-inductive bisimilarity invariants bisimulation

\subsection{Assertions as non-simulation certificates}
\label{sec:nonsim}

Hennessy and Milner \cite{hennessy85jacma} provided a characterization
of bisimulation in terms of an assertion language over modal operators
$[a]$ and $\diam{a}$.  The characterization states that two processes
are bisimilar if and only if they satisfy the same assertion
formulas.  Thus, if $p$ and $q$ are not bisimilar, there is some
assertion formula $A$ which is true for $p$ and not for $q$.
Formally, we write $p\models{}A$ and $q\not\models{}A$.

It is possible to use such assertion formulas directly as proof
certificates in the simpler and related problem of the absence of
simulation, \ie for theorems of the form ${\vdash\neg\ssim{p}{q}}$.
In that case,
the assertion language needs only the diamond modality $\diam\cdot$ as
well as the conjunction.  More formally, let \act be a set of actions.
The restricted set of assertions over \act is given by the recurrence
$%\[
  A:=\bigwedge_{i\in{}I}\diam{a_i}A_i
$%\]
, where $I$ is a finite set and $a_i\in\act$;
that is, we have a strict alternation of (indexed) conjunctions and
the diamond modality.
The statement $p\models\bigwedge_{i\in{}I}\diam{a_i}A_i$ means that,
for every $i\in{}I$, there exists a $q_i$ such that
$\one{p}{a_i}{q_i}$ and $q_i\models{}A_i$.
We shall choose to write $\atrue$ for empty
conjunctions and we can drop $\bigwedge_{i\in I}$ when $I$ is a
singleton.  Thus, $\diam{a}\atrue$ stands for
$\bigwedge_{i\in\{\bullet\}}\diam{a}
 \bigwedge_{j\in\{\}}\diam{b_{i,j}}A_{i,j}$.

Some of the clerks and experts
needed for this interpretation of an assertion as a certificate are
listed below; the rest of the definition can be taken from the
\asynccert{} constructor.
\begin{align*}
  \forall(a_i)_i\forall(A_i)_i\forall{}j.\,& \decidele(\bigwedge\strut_i\diam{a_i}A_i,\diam{a_j}A_j). & \forall{}a\forall{}A.\,& \foralle(\diam{a}A,A,a). \\
  \forall{}A.\,                            & \nimpe(A,\synccert(\stopcert),A).                        & \forall{}T\forall{}A.\,& \foralle(A,A,T). \\
  \forall{}a\forall A.\,                   & \nule(\diam{a}A,\diam{a}A).                              & \forall{}A.\,          & \releasele(A,A).
\end{align*}

\begin{example}
  In \Cref{fig:LTS1}, the process $(6)$ is not simulated by the process
  $(1)$: if $\Xi$ is the assertion formula
  $\diam{a}(\diam{b}\atrue\wedge\diam{c}\atrue)$, then $6\models\Xi$
  but $1\not\models\Xi$.
\end{example}

%%  LocalWords:  bisimilarity bisimulation bisimilar modality

\subsection{Assertions as non-bisimilarity certificates}
\label{sec:nonbisim}

It is possible to extend the FPC described in \Cref{sec:nonsim} to
account for the absence of bisimulation in addition to the absence of
simulation.  As bisimilarity is finer than similarity, this will
require a richer class of assertion formulas.
The fact that it is a symmetric relation suggests that assertions
should contain negations.

We could use full Hennessy-Milner logic (\ie any arbitrary mix of
$\diam\cdot$, $[\cdot]$, $\vee$ and $\wedge$ or, equivalently,
$\diam\cdot$, $\wedge$ and $\neg$), but instead we choose the smaller
but equivalent set of assertions defined by the following recurrence.
\begin{align*}
  A &:=\bigwedge_{i\in{}I}B_i \\
  B &:=\diam{a_i}A_i\;|\;\neg(\diam{a_i}A_i)
\end{align*}
It can be shown that this set characterizes the same relation as full
Hennessy-Milner logic.
% \begin{proof}
% under one modal and above the modals under it, we can use
% distributivity to get <>(\/(/\)) or [](/\(\/)), and since \/ (resp.
% /\) commutes with <> (resp. []), we get
%   A := <a>/\_iA_i | [a]\/_iA_i
% ie
%   A := <a>/\_iA_i | not(<a>/\_iA_i)
% at the end, we might end up with
% disjunctions at the top-level, but they are useless for the
% characterization, hence the chosen set of assertions
% \end{proof}
The statement $p\models\bigwedge_{i\in{}I}B_i$ means that, for every
$i\in{}I$, $p\models{}B_i$; the statement $p\models\diam{a}A$ means
that there exists a $q$ such that $\one{p}{a}{q}$ and $q\models{}A$;
and the statement $p\models\neg(\diam{a}A)$ means that
$p\not\models\diam{a}A$.

Very little more is needed to extend the FPC to handle this
certificate.  We need to make sure that, in addition to certificates
with a top-level $\diam\cdot$, $\decidele$ and $\nule$ allow (and
propagate) certificates with a top-level $\neg\diam\cdot$.
We also need an expert to consume $\neg$,
and an expert to handle the additional $\nand$ connective (see the
definition of bisimilarity from \Cref{ex:lts}).  If we give these two
roles to the same new expert, namely $\nande$, the link between
reflexivity and negations in the assertions appears clearly.

The resulting set of clerks and experts for theorems of the form
$\vdash\neg\bisim{p}{q}$ is the following.
\begin{align*}
  \forall{}A.\,              & \storelc(A,A). &
  \forall(B_i)_i\forall{}j.\,& \decidele(\bigwedge\strut_iB_i,B_j). \\\rule{0pt}{3ex}
  \forall{}B.\,              & \nule(B,B). &
  \forall{}a\forall{}A.\,    & \nande(\phantom{\neg}\diam{a}A,\diam{a}A,1). \\
  \forall{}a\forall{}A.\,    & \foralle(\diam{a}A,A,a). &
  \forall{}a\forall{}A.\,    & \nande(\neg\diam{a}A,\diam{a}A,2). \\
  \forall{}T\forall{}A.\,    & \foralle(A,A,T). &
  \forall{}A.\,              & \nimpe(A,\synccert(\stopcert),A). \\\rule{0pt}{3ex}
  \forall{}A.\,              & \releasele(A,A). \\\rule{0pt}{3ex}
  \forall{}A.\,              & \existsc(A,\lambda{}x.\,A). &
  \forall{}A.\,              & \pandc(A,A). \\
  \forall{}A.\,              & \mule(A,A). &
  \forall{}A.\,              & \porc(A,A,A). \\
  \forall{}A.\,              & \eqsc(A,A).
\end{align*}
This FPC extension is conservative, in that it can still check a
certificate for non-simulation.

\begin{example}
  In \Cref{fig:LTS1}, the processes $(6)$ and $(10)$ are similar but not
  bisimilar: if $\Xi$ is the generalized assertion formula
  $\diam{a}\neg\diam{b}\atrue$, then $10\models\Xi$ but
  $6\not\models\Xi$.
\end{example}

\section{A reference proof checker}
\label{sec:reference}

The framework for foundational proof certificates described in~%
\cite{miller11cpp,chihani13cade} was based on proof theory without
fixed point definitions.  In that setting, a standard logic
programming language (in that case, $\lambda$Prolog
\cite{miller12proghol}) was an ideal prototyping language for
implementing and testing FPCs.  The FPCs described in this paper are
not so easily implemented in standard logic programming languages
since the unification of eigenvariables must be done alongside the
usual unification of ``logic variables'' that makes proof
reconstruction possible.  The implementation of $\lambda$Prolog, for
example, considers eigenvariables as constants during unification.

We have built a prototype proof checker for testing the FPCs described
in this paper using the Bedwyr extension to logic programming
\cite{Bedwyr,baelde07cade}.
That system, originally designed to tackle various kinds of model
checking problems, provides the necessary unification for logic and
eigenvariables along with backtracking search and support for
$\lambda$-terms, $\lambda$-conversion and higher-order pattern
unification.

One could have imagined implementing the non-augmented proof system
$\muF_0$ directly and in a sense, this is already done by Bedwyr
itself.  For
example, if $(\mu{}B\,\t)$ is a purely positive fixed point encoding a
Prolog predicate, when the system is given the sequent
$\syncR{}{(\mu{}B\,\t)}{}$, it would emulate the Prolog search.
Similarly, if it is given the sequent
$\async{}{}{(\mu{}B\,\t)}{}{}{}$, it would emulate a finite failing
proof search.   But, as anyone familiar with Prolog-style depth-first
search knows, such proof search is limited in its effectiveness.  For
example, if one is attempting to prove that there is or is not a path
between two points, a cycle in the underlying graph can make the
search non-terminating.
Bedwyr handles this with a loop-detection mechanism that can be
embedded in the rules from \Cref{fig:muFa1}, making it a partial
implementation of \muF.

However, our goal with the \muF{} proof system is not to use it by
itself, but together with clerks and experts, as the engine (as a
``kernel'') for checking already existing proof evidence.  Since the
logic of the existing Bedwyr system has no native support for proof
objects, we implemented \muFa{} as an ``object logic'', without using
some native features such as loop-detection.
The Bedwyr specification files that we use (available directly
at \url{http://slimmer.gforge.inria.fr/bedwyr/pcmc/}, or
from the authors' homepages) are rather direct translations of the
inference rules in \Cref{fig:muFa0,fig:muFa1} as well as of the various FPCs
listed in the previous few sections.  It has thus been easy for us to
experiment and test FPCs.

While we have found the Bedwyr system to be useful for prototyping a
proof checker, our proposal for FPC is not tied to any one particular
implementation.  Instead, the framework is defined using inference
rules (such as found in \Cref{fig:muFa0,fig:muFa1}).  Any system that
can implement the logical principles required by such inference rules
can be used as a proof checking FPC kernel.

%%  LocalWords:  foundational Prolog FPCs eigenvariables Bedwyr FPC
%%  LocalWords:  Bedwyr eigenvariables Prolog sequent FPC

\section{Conclusion}
\label{sec:conclusion}

We have taken the basic structure of \emph{foundational proof
  certificates} that had been developed elsewhere for first-order
logic and described how it could be imposed on a logic based on fixed
points.  The resulting logic is much richer (think of the difference
between first-order logic and first-order arithmetic) and additional
logic principles need to be accounted for in the description of proof
certificates.

In the areas of model checking that we have discussed, proof evidence
is often taken to be, say, a path through a graph, a set of pairs of
nodes (satisfying certain closure conditions), or a Hennessy-Milner
logic assertion formula.  We have illustrated how each of these
familiar objects can be easily transformed into hints to guide a proof
checker though the construction of a detailed and complete sequent
calculus proof.  The architecture of focused proof systems and the
clerk and expert predicates allow this conceptual gap (between
familiar proof evidence and sequent calculus proofs) to be bridged in a
flexible and natural fashion.

We have also provided a novel look at the proof theory foundations of
model checking systems by basing our entire project on the \mumall{}
variant of linear logic and on the notion of \emph{\switchable{}
formulas}.
This latter notion seems to provide an interesting demarcation between
the logically simpler notion of model checking and the more general
notion of (inductive and co-inductive) deduction.

\smallskip\noindent{\bf Acknowledgment.}
We thank the reviewers for their detailed and useful comments on an
earlier draft of this paper.
This work has been funded by the ERC Advanced Grant Proof\kern 0.6pt
Cert.

%%  LocalWords:  co-inductive

\newpage
\bibliographystyle{eptcs}
%\bibliography{../references/master}

\newpage
\appendix
\section[On the augmented focusing proof system muFa]{On the augmented focusing proof system \muFa}
\label{sec:muFa}

It should be noted that, although the system presents two left rules
for the connective $=$, one with the clerk $\eqsc$ for success and one
with the clerk $\eqfc$ for failure, the implementation is usually
expected to have one single unification facility that, given an
equation, will or will not succeed, and which is tied to one single
clerk.
If the unification fails, the rule succeeds immediately without
generating a premise certificate $\Xi_1$, and the constraint on the
conclusion certificate $\Xi_0$ is actually the same as for success.
Hence the clerk $\eqfc$ can be defined as an existential closure of
$\eqsc$.  Likewise, $\neqsc$ can be defined in terms of $\neqfc$.
\begin{align*}
  \eqfc(\cdot)  & \equiv(\exists\Xi_1.\,\eqsc(\cdot,\Xi_1)) &
  \neqsc(\cdot) & \equiv(\exists\Xi_1.\,\neqfc(\cdot,\Xi_1))
\end{align*}

It is also possible to remove the truth and falsity connectives, as we
expect to have the equivalences
\begin{align*}
  \ptrue  & \equiv(a=a)      & \pfalse & \equiv(a=b) \\
  \nfalse & \equiv(a\neq{}a) & \ntrue  & \equiv(a\neq{}b)
\end{align*}
if $a$ and $b$ are distinct constants, hence the following:
\begin{align*}
  \ptruee  & \equiv\eqse  & \ptruec  & \equiv\eqsc  & \pfalsec & \equiv\eqfc \\
  \nfalsee & \equiv\neqfe & \nfalsec & \equiv\neqfc & \ntruec  & \equiv\neqsc
\end{align*}

Last, it is customary to leave clerks and experts out of rules with no
premises (\ie $\ptruee$, $\eqse$, $\nfalsee$, $\neqfe$, $\pfalsec$,
$\eqfc$, $\ntruec$ and $\neqfc$).  This has the same effect as setting
them to be always true.

The system presented in \Cref{fig:muFa0,fig:muFa1} does not have these
simplifications, but the Bedwyr-based implementation does.

\section{A simple example of a $\muF$ proof}
\label{sec:example}

The following proof can be seen as the justification that
$\{1,3\}\subseteq\{1,2,3\}$.  In particular, encode these two sets
as the predicates (i.e., abstractions over formula):
\[\lambda x[x = 1\por x = 3] \hbox{\quad and \quad}
  \lambda x[x = 1\por x = 2 \por x = 3].\]
The sequent calculus proof of inclusion can then be written as the
following focused proof.
\[
\infer{\async[]{}{}{}{\forall x. [x = 1\por x = 3]\nimp[x = 1\por x = 2 \por x = 3]}{}{}}
{\infer{\async[]{}{}{}{[x = 1\por x = 3]\nimp[x = 1\por x = 2 \por x = 3]}{}{}}
{\infer{\async[]{}{}{x = 1\por x = 3}{x = 1\por x = 2 \por x = 3}{}{}}
       {\infer{\async[]{}{}{x = 1}{x = 1\por x = 2 \por x = 3}{}{}}
       {\infer{\async[]{}{}{}{1 = 1\por 1 = 2 \por 1 = 3}{}{}}
       {\infer{\async[]{}{}{}{}{1 = 1\por 1 = 2 \por 1 = 3}{}}
       {\infer{\syncR[]{}{1 = 1\por 1 = 2 \por 1 = 3}{}}
       {\infer{\syncR[]{}{1 = 1}{}}{}}}}}
        \ \
       \infer{\async[]{}{}{x = 3}{x = 1\por x = 2 \por x = 3}{}{}}
       {\infer{\async[]{}{}{}{3 = 1\por 3 = 2 \por 3 = 3}{}{}}
       {\infer{\async[]{}{}{}{}{3 = 1\por 3 = 2 \por 3 = 3}{}}
       {\infer{\syncR[]{}{3 = 1\por 3 = 2 \por 3 = 3}{}}
       {\infer{\syncR[]{}{3 = 3}{}}{}}}}}
}}}
\]

\end{document}